# PROACTIVE BOTTLENECK PERFORMANCE ANALYSIS IN PARALLEL COMPUTING USING OPENMP

Vibha Rajput
Computer Science and Engineering
M.Tech.2nd Sem (CSE)
Indraprastha Engineering College. M. T.U
Noida, U.P., India

Alok Katiyar
Computer Science and Engineering
Faculty of Computer Science Department,
Indraprastha Engineering College,
Ghaziabad, U.P., India

*Abstract*—The aim of parallel computing is to increase an application's performance by executing the application on multiple processors. OpenMP is an API that supports multi-platform shared memory programming model and shared-memory programs are typically executed by multiple threads. The use of multi threading can enhance the performance of application but its excessive use can degrade the performance. This paper describes a novel approach to avoid bottlenecks in application and provide some techniques to improve performance in OpenMP application. This paper analyzes bottleneck performance as bottleneck inhibits performance. Performance of multi threaded applications is limited by a variety of bottlenecks, e.g. critical sections, barriers and so on. This paper provides some tips how to avoid performance bottleneck problems. This paper focuses on how to reduce overheads and overall execution time to get better performance of application.

## I. INTRODUCTION

Parallel computing has been considered to be "the high level of computing", and has been used to model difficult problems in many areas of science and engineering such as Computer Science, Mathematics, Defense, and so on [16]. Parallel Computing is simultaneous use of more than one CPU or processor core to execute a program or multiple computational threads. Ideally, parallel processing makes programs run faster because there are more engines (CPUs or cores) running it[17].In practice, developing a parallel application involves partitioning workload into tasks and mapping of tasks into workers. Two major models [1] for parallel programming are the message-passing model and the shared address space model. In the message-passing model, each processor has private memory and communicates data to other processors by a message. The Message Passing Interface (MPI) [15] is a de facto standard interface of this model. In the shared address space model, all processors share a memory and communicate data through the access to the shared memory .The OpenMP [12] (Open-multiprocessing) is a de facto standard interface of this model. It is mainly used for a shared memory multiprocessor (SMP) machine. The programming model of OpenMP is based on cooperating threads running simultaneously on multiple processors or cores. Threads are created and destroyed in a fork–join pattern. All OpenMP programs begin as a single process-master thread. The master thread executes sequentially until the first parallel region construct is encountered.

This API is simple and flexible for developing parallel applications for platforms ranging from desktop to supercomputer. It is comprises a set of compiler directives, runtime routines and environment variables to specify shared-memory parallelism in FORTRAN and C/C++ programs. An OpenMP directive is a specially formatted comment or pragma that generally applies to the executable code immediately following it in the program. A directive or OpenMP routine generally affects only those threads that encounter it.

> #pragma omp directive-name [clause [[,] clause]
>     . . . ] new-line

Figure 1. General form of an OpenMP directive for C/C++ programs.

The directive-name is a specific keyword, for example parallel, that defines and controls the action(s) taken. The clauses can be used to further specify the behavior.

Although creating an OpenMP program [10] can be easy, sometimes simply inserting directives is not enough. The resulting code may not deliver the expected level of performance, and it may not be obvious how to remedy the situation. This paper suggested some techniques that may help to improve performance. Identifying performance bottlenecks is important for application developers to produce high performance application software. Effective performance analysis becomes essential for application developers to diagnose the code behavior and provide optimizations. Performance bottlenecks are places in application that prevent the application from running as fast as it should. OpenMP supports multi threaded application and the performances of multi-threaded applications are limited by a variety of bottlenecks such as critical sections and barriers. These bottlenecks serialize execution, waste valuable execution cycles, and limit scalability of applications [2].

This paper analyzes performance of an application with various no. of threads and identifying bottlenecks from that. Researchers use an ompP profiler to find out overheads and to reduce overall execution time. This paper proposes some set of rules that are useful for developers for writing efficient program .This paper is organized as follows. In section 2, paper present motivation of our study. In section 3, paper present methodology of proposed work. In section 4, paper focus on performance enhancement and section 5, paper present experimental results and analysis of their application. Researchers implement matrix multiplication application with various matrix sizes and with various numbers of threads.

## II. MOTIVATION

Researchers can define bottleneck problem in terms of programming approach is the portion of code that inhibits performance of overall application. Bottlenecks [2] cause thread serialization. Thus, a parallel program that spends a significant portion of its execution in bottlenecks can lose some or even all of the speedup that could be expected from parallelization. In OpenMP a single thread or multiple threads that need to reach a synchronization point before other threads can make progress. Thus an efficient implementation of synchronization directives is crucial to overall performance.

### A. Synchronize Threads

Synchronizing, or coordinating the actions of threads is sometimes necessary in order to ensure the proper ordering of their access to shared data and to prevent data corruption. OpenMP uses synchronization directives to synchronize threads. The synchronization directive is essential to overall performance since synchronization operations are very expensive and inhibit concurrency in applications. Critical construct and barrier construct are two most common forms of synchronization directive in OpenMP. These constructs may arise performance bottlenecks.

#### 1) Critical Section

Only one thread can execute a critical section at a given time, and any other thread wanting to execute the critical section must wait That is, there is only one executer and possibly multiple waiters Fig.2 shows the execution of four threads (T1-T4) that run a simple loop that consists of a critical section C1 and a non-critical section N. Threads can execute in parallel until time 30, at which T3 acquires the lock and prevents T1 and T4 from doing so. Then, T1 and T4 become idle, waiting on T3. Accelerating the execution of C1 on T3 during this time would not only accelerate T3 and reduce the total execution time (because this instance of code segment C1 is on the critical path), but also minimize the wait for both T1 and T4. However, accelerating the same C1 on T2 between times 20 and 30 would not help because all the other threads are running useful work and no thread is waiting for the lock [2].

#### 2) Barriers

Threads that reach a barrier must wait until all threads reach the barrier. There can be multiple executers and multiple waiters. Fig. 3 shows the execution of four threads reaching a barrier. T1 and T4 reach the barrier at time 30 and start waiting on T2 and T3. T2 reaches the barrier at time 40, leaving T3 as the only running thread until time 60, when T3 reaches the barrier and all four threads can continue. Therefore, every cycle saved by accelerating T3 gets subtracted from the total execution time, up to the difference between the arrivals of T2 and T3 at the barrier [2].

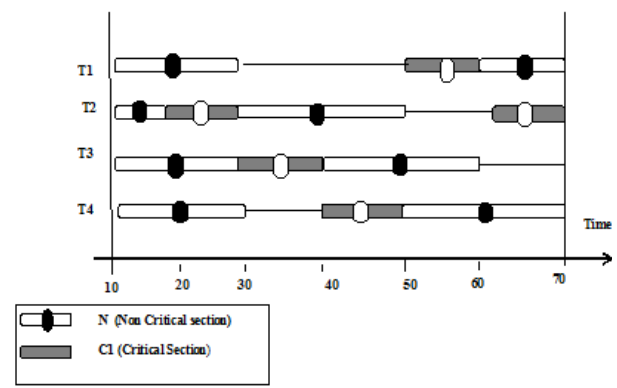

Figure 2.   Critical Section

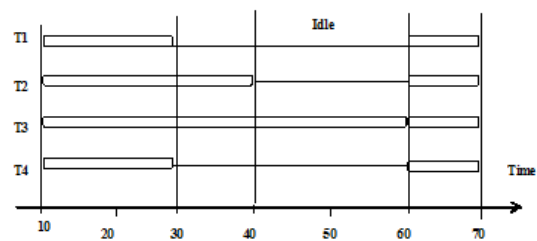

Figure 3.   Barrier

### B. Causes of poor performance

Synchronization, thread management, load imbalance, memory access are the main factors that influence the performance of OpenMP application. Synchronization directives [8] are used to synchronize the threads; all threads are created at the beginning of execution and destroyed at the end of execution. Suspending and waking up the slaves are the main operations of thread management because of the fork and join programming model. Overheads [6] may arise at any operation of thread management. Proper thread management is essential for parallel programming and it gives higher performance in application.

In memory Access [10] when a program is compiled, the compiler will arrange for its data objects to be stored in main memory; they will be transferred to cache when needed. If a value required for a computation is not already in a cache (researchers call this a cache "miss"), it must be retrieved from higher levels of the memory hierarchy, a process that can be quite expensive. A major goal is to organize data accesses so that values are used as often as possible while they are still in cache. Bad memory access causes the poor utilization of memory system.

### III. METHODOLOGY OF PROPOSED WORK

*A. Performance measurement*

In this paper researchers have implemented matrix multiplication and measured performance using UNIX command
**$ >time . /matrix_multiplication.**
Below table 1 shows execution time of 1500*1500 square matrix multiplication with no. of threads.

Where **real time** is wall clock time, it is the time from start to finish of the program. **User time** is the amount of CPU time spent in user-mode code (outside any operating system services) within the process. This is only actual CPU time used in executing the process. **System time** is the time spent on operating system services, such as input/output routines.

| no. of thread | 2 | 4 | 8 |
|---|---|---|---|
| real time | 122.90 sec | 123.33 sec | 121.27 sec |
| user time | 121.42 sec | 122.01 sec | 120.54 sec |
| system time | 0.08 sec | 0.04 sec | 0.04 sec |

*B. Performance analysis*

Researchers have analyzed performance of matrix multiplication code using ompP tool. OmpP [4, 5] is a profiling tool for OpenMP applications written in C/C++ or FORTRAN. OmpP's profiling report becomes available immediately after program termination in a human-readable format. OmpP supports the measurement of hardware performance counters using PAPI [13] and supports several advanced productivity features such as overhead analysis. Four overhead categories are defined in ompP-

*a) Imbalance:* Waiting time incurred due to an imbalanced amount of work in a worksharing or parallel region.

*b) Synchronization:* Overhead that arises due to threads having to synchronize their activity, such as barrier call.

*c) Limited Parallelism*: Idle threads due to not enough parallelism being exposed by the program.

*d) Thread management*: Overhead for the creation and destruction of threads, and for signaling critical sections or locks as available.

Researchers have used following command to run our application with ompP profiler to analyze performance.
**$> kinst-ompp gcc -fopenmp matrix_multiplication -o matrix_multiplication**

After program termination researchers obtained a profiling report. According to this report our application with 2 threads have 2 parallel region, 4 parallel loops and 3 barrier construct. Total parallel coverage is 122.90 sec (100.00%).

Wall clock time of first parallel region is 0.09 sec. and second parallel region is 122.81 sec. Total overheads of parallel region first are 21.03 % and parallel region second are 0.34 %.

The above information shows that our application has more overheads [6] and poor performance. When performance is poor, it is often because some basic programming rules have not been adhered to. This paper provides some guidelines on how to write an efficient program to enhance the performance and also gives some tips to avoid common performance problems.

### IV. HOW TO IMPROVE PERFORMANCE

*A. Memory Access Patterns*

A modern memory system is organized as a hierarchy, where the largest, and also slowest, part of memory is known as main memory. Main memory is organized into pages. The memory levels closer to the processor are successively smaller and faster and are collectively known as cache. When a program is compiled, the compiler will arrange for its data objects to be stored in main memory; they will be transferred to cache when needed. If a value required for a computation is not already in a cache (researchers call this a cache "miss"), it must be retrieved from higher levels of the memory hierarchy, a process that can be quite expensive. Program data is brought into cache in chunks called blocks, each block will occupy a line of cache. Data that is already in cache may need to be removed, or "evicted", to make space for a new block of data.

A major goal is to organize data accesses so that values are used as often as possible while they are still in cache. The most common strategies for doing so are based on the fact that programming languages (including both FORTRAN and C) typically specify that the elements of arrays be stored contiguously in memory. Thus, if an array element is fetched into cache, "nearby" elements of the array will be in the same cache block and will be fetched as part of the same transaction. If a computation that uses any of these values can be performed while they are still in cache, it will be beneficial for performance [10].

*1) Good Memory Access*

When data has been brought into the cache(s), all the elements of the line are used before the next line is referenced. This type of access pattern is often referred to as "row wise storage". For good performance, should access the elements of the array row by row not column by column.

*2) Bad Memory Access*

When data has been brought into the cache(s), all the elements of the column are used before next row is referenced. This means that for an m x n array a, element a (1, 1) is followed by element a (2,1). The last element a (m, 1) of the first column in followed by a(1,2), etc. This is called "column wise storage."

TABLE I.  EXAMPLE OF MEMORY ACCESS

| Good memory access | Bad memory access |
|---|---|
| for (int i=0; i<n; i++)<br>   for (int j=0; j<n; j++)<br>     sum += a[i][j]; | for (int j=0; j<n; j++)<br>   for (int i=0; i<n; i++)<br>     sum += a[i][j]; |

*B.   Loop Optimizations*

Loop optimization [14] is the process of the increasing execution speed and reducing the overheads associated of loops. It plays an important role in improving cache performance and making effective use of parallel processing capabilities.

*1) Loop interchange (or loop exchange)*

These optimizations exchange inner loops with outer loops. This is one way in which loops can be restructured, or transformed, to improve performance.

*2) Loop unrolling*

This is a powerful technique to effectively reduce the overheads of loop execution (caused by the increment of the loop variable, test for completion). It has other benefits, too. Loop unrolling can help to improve cache line utilization by improving data reuse.

TABLE II.  EXAMPLE OF LOOP UNROLLING

| Normal loop | After loop unrolling |
|---|---|
| int i;<br>for(i=0;i<100;i++)<br>{<br>   add(i);<br>} | int i;<br>for(i=0;i<100;i+=4)<br>{<br>   add(i);<br>   add(i+1);<br>   add(i+2);<br>   add(i+3);<br>   add(i+4);<br>} |

*3)   Loop fusion/combining*

Another technique which attempts to reduce loop overhead When two adjacent loops would iterate the same number of times (whether or not that number is known at compile time), their bodies can be combined as long as they make no reference to each other's data. Loop fusion does not always improve run-time speed. Table 4 shows the example of loop fusion.

*4) Loop fission*

Loop fission attempts to break a loop into multiple loops over the same index range but each taking only a part of the loop's body.

TABLE III.  EXAMPLE OF LOOP FUSION

| Normal loop | After loop fusion |
|---|---|
| int i, a[100], b[100];<br>   for (i = 0; i < 100; i++)<br>     a[i] = 1;<br>   for (i = 0; i < 100; i++)<br>     b[i] = 2; | int i, a[100], b[100];<br>   for (i = 0; i < 100; i++)<br>   {<br>     a[i] = 1;<br>     b[i] = 2;<br>   } |

Above mentioned techniques are used for both sequential and parallel programming. Now researchers discuss on performance improvement factors for OpenMP applications. Researchers have briefly discussed synchronization, thread Management, task scheduling and memory access these are key factors that influence performance of OpenMP.

*C.   Synchronization*

If threads perform different amounts of work in a work-shared region, the faster threads have to wait at the barrier for the slower ones to reach that point. When threads are inactive at a barrier, they are not contributing to the work of the program. This is load imbalance overhead.

When threads typically waste time waiting for access to a critical region. If they are unable to perform useful work during that time, they remain idle. This is synchronization overhead [3].

To improve performance researchers should reduce these overheads by optimizing barrier use and avoid large critical region.

*1) Optimize Barrier Use*

No matter how efficiently barriers are implemented, they are expensive operations. The 'nowait' clause makes it easy to eliminate the barrier that is implied on several constructs.

*2) Avoid Large Critical Regions*

A critical region is used to ensure that no two threads execute a piece of code simultaneously. The more code contained in the critical region, however, the greater the likelihood that threads have to wait to enter it, and the longer the potential wait times. Therefore the programmer should minimize the amount of code enclosed within a critical region. If a critical region is very large, program performance might be poor.

*3) Avoid the Ordered Construct*

This is also synchronization construct. Ordered construct ensures that the corresponding block of code within a parallel loop is executed in the order of the loop iterations.

It is expensive to implement. The runtime system has to keep track which iterations have finished and possibly keep threads in a wait state until their results are needed. This inevitably slows program execution.

```
#pragma omp parallel shared (a) private (i)
{
for (i=0; i<dim; i++)
{
    a[i] = i+2;

    if (a [i]>10)
            smallwork (a[i])
    else
            bigwork (a[i])
}
```

Figure 4.   Example of Parallel loop with an uneven load

### D.   Scheduling Loops to Balance the Load

The manner in which iterations of a parallel loop are assigned to threads is called the loop's schedule. Using the default schedule [3] on most implementations, each thread executing a parallel loop performs about as much iteration as any other thread. When each iteration has approximately the same amount of work this causes threads to carry about the same amount of load and to finish the loop at about the same time. Generally, when the load is balanced fairly equally among threads, a loop runs faster than when the load is unbalanced.

Unfortunately it is often the case that different iterations have different amounts of work. Consider the code in Fig. 4. Each iteration of the loop may invoke either one of the subroutines smallwork or bigwork. Depending on the loop instance, therefore, the amount of work per iteration may vary in a regular way with the iteration number (say, increasing or decreasing linearly), or it may vary in an irregular or even random way. If the load in such a loop is unbalanced, there will be synchronization delays at some later point in the program, as faster threads wait for slower ones to catch up. As a result the total execution time of the program will increase [9]. By changing the schedule of a load-unbalanced parallel loop, it is possible to reduce these synchronization delays and thereby speed up the program. A schedule is specified by a schedule clause on the for loop directive.

The schedule [10] clause specifies how the iterations of the loop are assigned to the threads in the team. The syntax is **schedule (kind [, chunk size])**. Where chunk size must be a scalar integer value. The static schedule works best for regular workloads. For a more dynamic work allocation scheme the dynamic or guided schedules may be more suitable.

TABLE IV.   DIFFERENT KINDS OF SCHEDULE

| Schedule | Description |
|---|---|
| Static [,chunk] | 1. Distribute iterations in blocks of size "chunk" over the threads in a round-robin fashion. 2. In absence of "chunk", each thread executes approx. N/P chunks for a loop of length N and P threads |
| Dynamic [,chunk] | 1. Fixed portions of work; size is controlled by the value of chunk 2. When a thread finishes, it starts on the next portion of work |
| Guided [,chunk] | Same dynamic behavior as "dynamic", but size of the portion of work decreases exponentially |
| Run time[,chunk] | Iteration scheduling scheme is set at runtime through environment variable OMP_SCHEDULE |

### E.   Thread Management

Thread management [7] represents the actions performed during the life cycle of a thread. These actions include thread creation, activities of thread and their deletion. Task schedule is an activity of thread. It shows the total number of tasks performed by every thread.

OpenMP is based on fork and join concept of thread that means all programs begin as a single process-master thread. The master thread then creates a team of parallel threads. When the team threads complete the statements in the parallel region construct, they synchronize and terminate, leaving only the master thread.

### F.   Other Performance Factors

#### 1) Maximize Parallel Region

Indiscriminate use of parallel regions may give rise to suboptimal performance. Large parallel regions offer more opportunities for using data in cache and provide a bigger context for other compiler optimizations. Therefore it is worthwhile to minimize the number of parallel regions. Each parallelized loop adds to the parallel overhead and has an implied barrier that cannot be omitted. For example, if researchers have multiple parallel loops, researchers must choose whether to encapsulate each loop in an individual parallel region or to create one parallel region encompassing all of them.

#### 2) Avoid Parallel Region in Inner Loops

Parallel region embedded in loop nest shown in figure 2, the overheads of the **#pragma omp parallel for** construct are incurred n2 times. A more efficient solution is to split **#pragma omp parallel for** construct into its constituent directives and the **#pragma omp parallel** has been moved to enclose the entire loop nest. The **#pragma omp** for remains at the inner loop level.

TABLE V.                  MAXIMIZE PARALLEL REGION

| Multiple parallel regions | Single combined parallel region |
|---|---|
| #pragma omp parallel for<br>for (.....)<br>{<br>}<br>#pragma omp parallel for<br>for (.....)<br>{<br>}<br>..........<br>#pragma omp parallel for<br>for (.....)<br>{<br>} | #pragma omp parallel<br>{<br>#pragma omp for<br>{ ...... }<br>#pragma omp for<br>{ ...... }<br>..........<br>#pragma omp for<br>{ ...... }<br>} |

TABLE VI.                 AVOID PARALLEL REGION IN INNER LOOP

| Parallel region embedded in loop nest | Parallel region moved outside of the loop nest |
|---|---|
| for (i=0; i<n; i++)<br>for (j=0; j<n; j++)<br>#pragma omp parallel for<br>for (k=0; k<n; k++)<br>{ .........} | #pragma omp parallel<br>for (i=0; i<n; i++)<br>for (j=0; j<n; j++)<br>#pragma omp for<br>for (k=0; k<n; k++)<br>{ .........} |

## V.     EXPERIMENTS

### A.    Experimental Environment

Our experimental platform is Ubuntu -12.10 (a Linux version) with 2.16 GHz Intel Pentium dual core processor and 1.9 GB main memory. Researchers used GNU gcc compiler 4.7.2 with optimization level 3(-o3) option and OpenMP 3.1 API. Researchers have implemented matrix_multiplication application in OpenMP. Researchers analyzed performance of application with ompP 0.8.0 profiler.

Researchers implemented matrix_ multiplication and got an error of "segmentation fault" while running program of matrix size of 1000*1000 or beyond. So researcher allocated memory dynamically for the matrix to get more free space to accept matrix sizes of very large inputs. This mechanism is necessary for our purposes because with large input matrix sizes it is clearer to analyze performance.

*1) Experimental Results*
Researchers ran our application with (1000*1000)*(1000*1000), (1500*1500)*(1500*1500), (2000*2000)*(2000*2000), (2500*2500)*(2500*2500) matrices and obtained results shown in Table 8.To analyze overheads and slow response time, researchers have used ompP profiler and found that our application has 2 parallel region, 4 parallel loops and 3 barrier construct and its parallel coverage is about to 100%. The overheads analysis of our application is shown in table 9.

Researchers have modified our application according to guidelines that mentioned in previous section. The results of our modified application are showing in following table 10.With the help of profiler researchers have found that modified application has 1 parallel region, 3 parallel loops, and has no barrier construct. The overhead analysis of modified application is showing in table 11.

TABLE VII.     TOTAL EXECUTION TIME OF MATRIX MULTIPLICATION

| Matrix Input Size | Total Run Time | | |
|---|---|---|---|
| (n*n)*(n*n) | 2 Threads | 4 Threads | 8 Threads |
| (1000*1000)(1000*1000) | 35.72 Second | 35.33 Second | 34.99 Second |
| (1500*1500)*(1500*1500) | 122.90 Second | 123.33 Second | 121.27 Second |
| (2000*2000)*(2000*2000) | 295.00 Second | 308.08 Second | 294.77 Second |
| (2500*2500)*(2500*2500) | 694.13 Second | 660.83 Second | 670.25 Second |

TABLE VIII.     OVERHEAD ANALYSIS

| With 2 Threads | Ovhds (%) = Synch (%) + Imbal (%) + Limpar (%) + Mgmt (%) |
|---|---|
| (1000*1000)(1000*1000) | 0.36 ( 0.51)   0.00 ( 0.00)   0.36 ( 0.51)   0.00 ( 0.00)   0.00 ( 0.00) |
| (1500*1500)(1500*1500) | 0.87 ( 0.35)   0.00 ( 0.00)   0.86 ( 0.35)   0.00 ( 0.00)   0.01 ( 0.00) |
| (2000*2000)(2000*2000) | 1.51 ( 0.26)   0.00 ( 0.00)   1.51 ( 0.26)   0.00 ( 0.00)   0.00 ( 0.00) |
| (2500*2500)(2500*2500) | 3.03 ( 0.22)   0.00 ( 0.00)   3.03 ( 0.22)   0.00 ( 0.00)   0.00 ( 0.00) |

TABLE IX.     TOTAL EXECUTION TIME OF MODIFIED MATRIX MULTIPLICATION

| Matrix Input Size | Total Run Time | | |
|---|---|---|---|
| (n*n)*(n*n) | 2 Threads | 4 Threads | 8 Threads |
| (1000*1000)(1000*1000) | 11.51 Second | 11.05 Second | 11.41 Second |
| (1500*1500)*(1500*1500) | 40.04 Second | 47.19 Second | 44.09 Second |
| (2000*2000)*(2000*2000) | 123.84 Second | 136.27 Second | 142.71 Second |
| (2500*2500)*(2500*2500) | 323.29 Second | 432.06 Second | 329.16 Second |

TABLE X.     OVERHEADS ANALYSIS

| With 2 Threads | Ovhds (%) = Synch (%) + Imbal (%) + Limpar (%) + Mgmt (%) |
|---|---|
| (1000*1000)(1000*1000) | 0.04 ( 0.16)   0.00 ( 0.00)   0.04 ( 0.16)   0.00 ( 0.00)   0.00 ( 0.01) |
| (1500*1500)(1500*1500) | 0.05 ( 0.06)   0.00 ( 0.00)   0.05 ( 0.06)   0.00 ( 0.00)   0.00 ( 0.00) |
| (2000*2000)(2000*2000) | 0.01 ( 0.00)   0.00 ( 0.00)   0.00 ( 0.00)   0.00 ( 0.00)   0.01 ( 0.00) |
| (2500*2500)(2500*2500) | 0.48 ( 0.07)   0.00 ( 0.00)   0.48 ( 0.07)   0.00 ( 0.00)   0.00 ( 0.00) |

*1) Analysis*

Researchers have done overhead analysis to reduce overheads from application, which is shown in table 9. In this table the **imbal**(Imbalance) category have overheads of 0.51%,0.35%,0.26%,0.22% with matrix sizes of [1000*1000],[1500*1500],[2000*2000],[2500*2500] respectively. This category shows that our application takes more waiting time due to an imbalanced amount of work in a worksharing or parallel region.

To reduce these overheads researchers have maximized parallel region and reduced waiting time for each thread with the help of loop scheduling, loop optimization and proper thread management. Apart from this researchers have reduced barrier constructs and avoided order construct. Table 10 shows total run time of modified application which has faster run time and better performance. Now **imbal** category of modified application with various matrix sizes are 0.16%, 0.06%,0.00%,0.48% that shown in Table 11.Researchers have drawn graphs fig.5&6 ,corresponding to performance of our applications

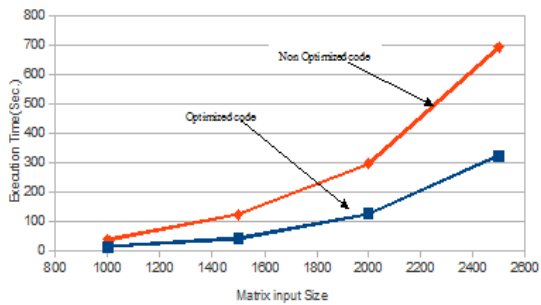

Figure 5.    Performance graph with 2 threads

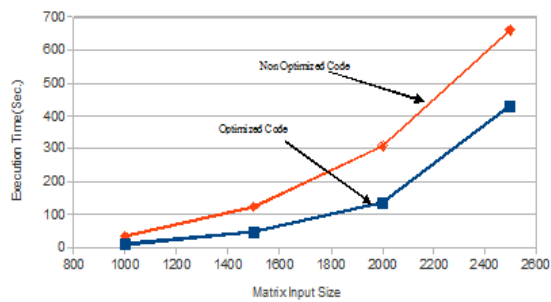

Figure 6.    Performance graph with 4 threads

## VI. CONCLUSION AND FUTURE WORK

In this paper, researchers identified bottleneck of our OpenMP applications. Researchers studied how to reduce bottleneck from application and learned how to improve performance of application using below techniques.

- Memory access pattern should be properly.
- Loop optimization should be used.
- Maximize parallel region.
- Avoid large critical region.
- Scheduling load should balance.

Researchers also learned some major factors that influence the performance such as synchronization, load imbalance, thread management and etc. In this paper researchers analyzed performance of matrix multiplication application with various no. of threads. To improve the performance of this application, researchers applied some strategies that mentioned in section 4 and finally researchers got better performance than previous.

The future enhancement of this work is highly laudable as parallelization using OpenMP is gaining popularity these days. This work will be carried out in the near future for various real time implementations like image processing, cloud computing and weather forecasting.

ACKNOWLEDGMENT

This research paper is made possible through the help and support from everyone, including: parents, teachers, family and friends.

First, I would like to thank Dr. P.N.Hrisheekesha, Director and Prof. O.P. Singhal, HOD, CSE Dept., Indraprastha Engineering College, for their support and encouragement.

Second, I would like to thanks to all our Lab colleague of Indraprastha Engineering College & staff members for extending a helping hand at every juncture of need.

Finally, I sincerely thank to my parents, family, and friends, who provide the advice and financial support. This research paper would not be possible without all of them.

.